\documentclass[%
 reprint,
 amsmath,amssymb,
 aps,
prb,
floatfix,
]{revtex4-2}

\usepackage{graphicx}
\usepackage{dcolumn}
\usepackage{bm}
\usepackage{physics}

\usepackage{subcaption}
\usepackage{tikz}
\usetikzlibrary{3d}
\usepackage{slashed}
\usetikzlibrary{calc,decorations.markings}

\newcommand{\Romannumeral}[1]{\MakeUppercase{\romannumeral #1}}
\newcommand{\vk}{\mathbf{k}}

\begin{document}


\title{Proximity effect of s-wave superconductor on inversion broken Weyl Semi-Metal}

\author{Robert Dawson}
\email{robert.dawson001@ucr.edu}
\author{Vivek Aji}
 \email{vivek.aji@ucr.edu}
\affiliation{
 Department of Physics \& Astronomy, University of California Riverside
}
\date{\today}


\begin{abstract}
    Inducing superconductivity in systems with unconventional band structures is a promising approach for realising unconventional superconductivity. Of particular interest are single interface or Josephson Junction architectures involving Weyl semimetals, which are predicted to host odd parity, potentially topological, superconducting states. These expectations rely crucially on the tunneling of electronic states at the interface between the two systems. In this study, we revisit the question of induced superconductivity in an inversion broken WSM via quantum tunneling, treating the interface as an effective potential barrier. We determine the conditions under which the gap function couples to the Weyl physics and its properties within the WSM. Our simulations show that the mismatch in the nature of the low energy electronic states leads to a rapid decay of the superconductivity within the semi-metal. 
\end{abstract}

\maketitle

\section{Introduction}

New quantum states of matter have been predicted to appear when materials with nontrivial band structure are placed in proximity with correlated phases such as superconductivity and magnetism. However, a key requirement is the ability of the electronic states on the two sides of the interface to admix efficiently. While symmetry considerations and effective models predict possible induced phases, a quantitative understanding can only be ascertained by a detailed modeling of the relevant physics at the interface. In this paper, we discuss the nature of superconductivity induced in a time reversal preserving Weyl semimetal (WSM) when placed in proximity to an s-wave superconductor within a continuum model. Motivated by the experimental observation in ion-irradiated NbAs microstructures \cite{doi:10.1126/sciadv.1602983}, our primary focus is on understanding the conditions under which superconductivity can be induced in a WSM. Unlike past approaches that either (a) rely on a tunneling model \cite{PhysRevB.90.195430,PhysRevB.102.014513,PhysRevB.100.035447,Meng2012WeylS,Bednik_2016,PhysRevB.92.035153,Burkov_2011} or (b) use effective field theory approaches starting from a phenomenological pairing model \cite{PhysRevB.89.014506,PhysRevLett.120.067003,PhysRevB.102.014513,PhysRevLett.100.096407,PhysRevB.79.161408,PhysRevB.86.214514,PhysRevLett.114.096804,PhysRevLett.113.046401,PhysRevB.93.184511}, we numerically solve the Bogoliubov DeGennes equation for the full superconductor-WSM system. The purpose of this approach is to establish the effectiveness of the proximity effect to induce superconductivity which faithfully accounts for the mismatch of material parameters such as location of low energy states in the Brillouin zone, Fermi velocity, and the symmetry of the wave-functions. 
\par This paper is organized as follows: In section \Romannumeral 2, we introduce models for the WSM and Superconducting materials, and establish our Nambu basis. In section \Romannumeral 3, we detail the numerical method we use to calculate the superconducting pairing amplitudes. In section \Romannumeral 4, we explore the results for (1) A Superconductor-WSM architecture, (2) A Josephson Junction architecture, and (3) An exploration of Superconductor-WSM architectures to identify key parameters. Finally, we discuss the implications of our findings in section \Romannumeral 5. 

\section{Model}

\begin{figure*}
    \centering
    \includegraphics[width=0.6\textwidth]{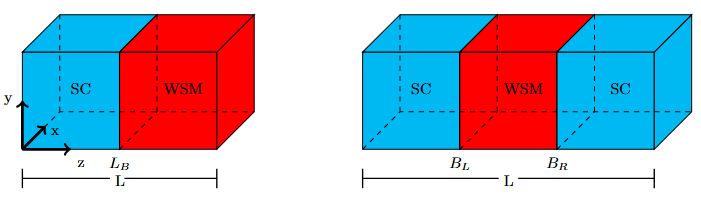}
     \caption{(Left) Device consisting of an S-Wave superconductor (SC) in contact with a WSM (WSM); the boundary is located at $L_B$. (Right) A Josephson Junction architecture with boundaries at $B_L$ and $B_R$.}
     \label{fig:Schematic}
\end{figure*}

Consider a device of length $L$ with a boundary at $z=L_B$ separating: (1) A Metallic Superconductor and (2) A WSM with broken inversion symmetry. The Hamiltonian for the WSM is \cite{Jang_2018}:
    \begin{widetext}
    \begin{equation}\label{eq:WSM_Base}
        H_{WSM} = \int\mathbf{d^3r}\sum_{ss'}\sum_{\sigma\sigma'}\Psi^\dagger_{s\sigma}(\mathbf{r})\bigg[v\sigma_xS_zP_x - v\sigma_yP_y + (m_zP^2_z - m)\sigma_z
         -\mu\sigma_0S_0\bigg]\Psi_{s'\sigma'}(\mathbf{r})
         \end{equation}
    \end{widetext}
    Here, $\Psi^\dagger_{s\sigma}$ ($\Psi_{s\sigma}$) is a creation (annihilation) operator for an electron with spin $s=\uparrow,\downarrow$ and orbital quantum number $\sigma=1,2$. The momentum operator is given by $\mathbf{P} = -i\vec{\nabla}$, $\mu$ is the chemical potential, and the Pauli matrices $\sigma_i$ ($S_i$) act in the orbital (spin) subspace, with $S_0$ and $\sigma_0$ their respective identities. Parity and Time-reversal operators are $P=\sigma_{z}$ and $T = \imath S_{y}\mathcal{K}$ where $\mathcal{K}$ performs complex conjugation. Of the four possible terms that break inversion symmetry, but preserve time reversal, two generate nodal rings while the other two generate Weyl nodes in either the $k_{x}-k_{z}$ or $k_{y}-k_{z}$ plane. Focusing on nodal phenomena the term that has Weyl nodes in the $k_{x}-k_{z}$ is given by
    
    \begin{align}
        \label{eq:Inv_Break}
        H_{IB} = \int\mathbf{d^3r}\sum_{ss'}\sum_{\sigma\sigma'}\Psi^\dagger_{s\sigma} (\alpha\sigma_x)\Psi_{s'\sigma'}(\mathbf{r})
    \end{align}

The distance between the nodes along the $k_{x}$ direction is given by $2\alpha$ which in principle can be determined from data on Weyl semi-metals. We treat it as a phenomenological parameter in our effective model. An important consideration in constructing a continuum model is to account for the discontinuity in the Hamiltonian along the z-direction. Our goal is to use a basis that spans the entire device, which requires the same degrees of freedom and power of the $P_{z}$ operator on both sides of the interface. To implement the numerical procedure in a Fourier basis, the model for the metallic region is also written in the two band basis. The parameters are chosen so that a quadratically dispersing spin degenerate band centred about the $\Gamma$ point intersects the Fermi surface:
    
    \begin{align}
        \label{eq:Metallic_Model}
        H_M &= \int_{\mathbf{r}}\sum_{ss'}\sum_{\sigma\sigma'}\Psi^\dagger_{s\sigma}(\mathbf{r})\bigg[(m_z\mathbf{P}^2+E_0) \sigma_zS_0\nonumber\\ &- \mu\sigma_0S_0\bigg]\Psi_{s'\sigma'}(\mathbf{r}) 
    \end{align}
    where $E_0$ is the band gap. Finally, we add to the metallic model a superconducting term:
    \begin{align}
        \label{eq:SC_Model}
        H_{SC} &= \int_{\mathbf{r}}\sum_{ss'}\sum_{\sigma\sigma'}(iS_y)_{ss'}\Delta_{\sigma\sigma'}(\mathbf{r})\Psi^\dagger_{s\sigma}(\mathbf{r})\Psi^\dagger_{s'\sigma'}(\mathbf{r})+ H.c.
    \end{align}
    The gap function $\Delta_{\sigma\sigma'}$ is given by:
    \begin{align}
        \label{eq:Orbital_Gap}
        \Delta_{\sigma\sigma'}(\mathbf{r}) &= g_{\sigma\sigma'}(\mathbf{r})F_{\sigma\sigma'}(\mathbf{r})
    \end{align}
    where $F_{\sigma\sigma'}(\mathbf{r})$ is the pairing amplitude. The interaction strength $g_{\sigma\sigma'}(\mathbf{r})$ is constant within the superconductor, and only nonzero for $\sigma=\sigma'=1$. Note that the $\sigma=1$ band is the only one that intersects the chemical potential. The advantage of this representation is that the Hamiltonian can be expanded in a basis of $\sin(k_{z}z)$ with the choice of $k_{z}$ determined by the boundary condition at $z=0$ and $z=L$. A self consistent solution accurately captures the interface at $z=L_{B}$ represented by a sharp discontinuity.
    
    \par We obtain the BdG Hamiltonian for the Bulk Model in the Nambu basis:
    \begin{align}
        \label{eq:Nambu Basis}
        \mathbf{\Psi}_{\mathbf{k} _\perp}(z) &= [\Psi_{\mathbf{k} _\perp,1,\uparrow},\Psi_{\mathbf{k} _\perp,1,\downarrow},\Psi_{\mathbf{k} _\perp,2,\uparrow},\nonumber\\ &\Psi_{\mathbf{k} _\perp,2,\downarrow},\Psi^\dagger_{-\mathbf{k} _\perp,1,\downarrow},\Psi^\dagger_{-\mathbf{k} _\perp,1,\uparrow},\Psi^\dagger_{-\mathbf{k} _\perp,2,\downarrow},\Psi^\dagger_{-\mathbf{k} _\perp,2,\uparrow}]^T
    \end{align}
   in the form
    \begin{align*}
        H &= \frac{1}{2}\int dz \int d^2\mathbf{k}_\perp \mathbf{\Psi}^\dagger_{\mathbf{k}_\perp}(z)\mathcal{H}_{BdG}(\mathbf{k}_\perp,z)\Psi_{\mathbf{k}_\perp}(z)
    \end{align*}
    with the BdG Hamiltonian:
    \begin{widetext}
    \begin{align}
        \mathcal{H}_{BdG}(\mathbf{k}_\perp,z) &= \sigma_z\tau_z[m_z\Theta(L_B - z)\mathbf{k}^2_\perp - m_z\partial^2_z + E_0(z)]
        + \tau_z[v(z)(k_x\sigma_xS_z - k_y\sigma_y) -  m(z)\sigma_z + \alpha(z)\sigma_x - \mu]\nonumber\\
        \label{eq:BdG_Hamil}
        &+ (iS_y)\Delta_{11}(z)\sigma_+\sigma_x\tau_+
        -(iS_y)\Delta_{11}(z)\sigma_+\sigma_x\tau_-
    \end{align}
    \end{widetext}
    where $\tau_i$ are the Pauli matrices in the particle-hole subspace, $\tau_{\pm} = (\tau_x \pm i\tau_y)/2$, and $\Delta_{11}$ is the gap function of the host superconductor. The parameters $E_0,m,v,$ and $\alpha$ have been replaced with piece-wise functions that are nonzero only within their respective regions. 
    \par With broken inversion symmetry and the introduction of the orbital quantum number $\sigma$, four nonzero pairing amplitudes are allowed:
    {\footnotesize
    \begin{align}
        \label{eq:11_Pairing}
        F_{11}(z) &= -\frac{1}{2}\int d^2\mathbf{k}_\perp\bigg[\expval{\Psi_{-\mathbf{k}_\perp,1,\downarrow}\Psi_{\mathbf{k}_\perp,1,\uparrow}} + \expval{\Psi_{\mathbf{k}_\perp,1,\downarrow}\Psi_{-\mathbf{k}_\perp,1,\uparrow}}\bigg]\\
        \label{eq:S_Pairing}
        F_T(z) &= -\frac{1}{2}\int d^2\mathbf{k}_\perp\bigg[\expval{\Psi_{-\mathbf{k}_\perp,1,\downarrow}\Psi_{\mathbf{k}_\perp,2,\uparrow}} + \expval{\Psi_{\mathbf{k}_\perp,2,\downarrow}\Psi_{-\mathbf{k}_\perp,1,\uparrow}}\bigg]\\
        \label{eq:T_Pairing}
        F_S(z) &= -\frac{1}{2}\int d^2\mathbf{k}_\perp\bigg[\expval{\Psi_{-\mathbf{k}_\perp,1,\downarrow}\Psi_{\mathbf{k}_\perp,2,\uparrow}} - \expval{\Psi_{\mathbf{k}_\perp,2,\downarrow}\Psi_{-\mathbf{k}_\perp,1,\uparrow}}\bigg]\\
        \label{eq:22_Pairing}
        F_{22}(z) &= -\frac{1}{2}\int d^2\mathbf{k}_\perp\bigg[\expval{\Psi_{-\mathbf{k}_\perp,2,\downarrow}\Psi_{\mathbf{k}_\perp,2,\uparrow}} + \expval{\Psi_{\mathbf{k}_\perp,2,\downarrow}\Psi_{-\mathbf{k}_\perp,2,\uparrow}}\bigg]
    \end{align}
    }
    Where $F_S$ and $F_T$ denote an orbital singlet (spin triplet) and orbital triplet (spin singlet) pairing, respectively. 

\section{Method}

We seek to self consistently calculate the pairing amplitude given by Eqs.(\ref{eq:11_Pairing}-\ref{eq:22_Pairing}). To this end, we solve the BdG equation:
    \begin{align}\mathcal{H}_{\mathbf{k}_\perp}(z)\Phi_{\alpha,\mathbf{k}_\perp}(z) &= E_{\alpha,\mathbf{k}_\perp}\Phi_{\alpha,\mathbf{k}_\perp}(z)
    \end{align}
    where the wave function $\Phi_{\alpha,\mathbf{k}_\perp}(z)$ is given by:
    \[\Phi_{\alpha,\mathbf{k}_\perp}(z) =
    \begin{pmatrix}
        u_{\alpha,\mathbf{k}_\perp,1,\uparrow(z)}\\
        u_{\alpha,\mathbf{k}_\perp,1,\downarrow(z)}\\
        u_{\alpha,\mathbf{k}_\perp,2,\uparrow(z)}\\
        u_{\alpha,\mathbf{k}_\perp,2,\downarrow(z)}\\
        v_{\alpha,-\mathbf{k}_\perp,1,\downarrow(z)}\\
        v_{\alpha,-\mathbf{k}_\perp,1,\uparrow(z)}\\
        v_{\alpha,-\mathbf{k}_\perp,2,\downarrow(z)}\\
        v_{\alpha,-\mathbf{k}_\perp,2,\uparrow(z)}
    \end{pmatrix}
    \]
    and is subject to the boundary condition $\Phi_{\alpha,\mathbf{k}_\perp}(0) = \Phi_{\alpha,\mathbf{k}_\perp}(L) = 0$. We follow the approach taken in Setiwan et al. \cite{PhysRevB.99.174511}, and take the particle (hole) wave functions $u_{\alpha,\mathbf{k_\perp},\sigma,s}(z)$ ($v_{\alpha,\mathbf{k_\perp},\sigma,s}(z)$) to be:
    \begin{align}
        u_{\alpha,\mathbf{k_\perp},\sigma,s}(z) &= \sqrt{\frac{2}{L}}\sum_{n=1}^{N}u^{(n)}_{\alpha,\mathbf{k_\perp},\sigma,s}\sin(k_nz)\\
        v_{\alpha,\mathbf{k_\perp},\sigma,s}(z) &= \sqrt{\frac{2}{L}}\sum_{n=1}^{N}v^{(n)}_{\alpha,\mathbf{k_\perp},\sigma,s}\sin(k_nz)
    \end{align}
    where $k_n = n\pi/L$. The cutoff mode $N$ is chosen such that $N = (k^w_FL/\pi)\sqrt{1 + \omega_D/(\mu + m)}$, with $k^w_F$ the Fermi-Momentum of the WSM, and $\omega_D$ the Debye frequency. Under these conditions, Eqs.(\ref{eq:11_Pairing}-\ref{eq:22_Pairing}) are simplified from a coupled differential equation into a finite Eigenvalue problem that can be solved numerically. To do so, we recast the Hamiltonian in the chosen Fourier basis:
    \begin{align}
        \bra{n}\mathcal{H}(\mathbf{k}_\perp)\ket{m} &= \frac{2}{L}\int_0^Ldz \sin(k_nz)\nonumber\\
        &\qquad\quad\times\mathcal{H}_{\mathbf{k}_\perp}(z)\sin(k_mz)
    \end{align}
    As mentioned previously, this approach requires the same degrees of freedom and power of the $P_{z}$ operator on both sides of the interface. To illustrate, imagine the coefficient attached to $P_z$ ($m_z$) is not constant throughout the device; it will have a value of $m_L$ in the SC and $m_R$ in the WSM. The associated off-diagonal matrix elements $h_{nm}$ are now given by:
    \begin{align*}
        h_{nm}&=\frac{2m_L}{L}\int_0^{L_B}dzk^2_m\sin(k_nz)\sin(k_mz)\\
        &+ \frac{2m_R}{L}\int_{L_B}^{L}dzk^2_m\sin(k_nz)\sin(k_mz)\\
        & \propto (m_L - m_R)k^2_m
    \end{align*}
    Since this term is not symmetric under an exchange of $n$ and $m$, the resulting Hamiltonian is not Hermitian. Clearly the Fourier expansion is not the best approach without a proper accounting of the interface, but the numerical approach is still valid if an appropriate basis can be chosen. In general, the Bogoliubov-De Gennes (BdG) equation is rearranged into the differential equation:
    \begin{align}\label{eq:GeneralDiff}
        \sum_n m_n(z)(\partial_z)^n\mathbf{\Phi}_{\alpha,\mathbf{k}_\perp} &= A(z)\mathbf{\Phi}_{\alpha,\mathbf{k}_\perp}
    \end{align}
    Where $A(z)$ is a matrix that is typically constant in either region. Eq.(\ref{eq:GeneralDiff}) need not be solved in its entirety; if a general solution for $\mathbf{\Phi}_{\alpha,\mathbf{k}_\perp}$ can be found as a series with some set of basis functions, then this numerical approach can be implemented using those basis functions. 
    \par We are now equipped to self consistently calculate the pairing amplitudes given by Eqs.(\ref{eq:11_Pairing}-\ref{eq:22_Pairing}):
    \begin{align}\label{eq:Pairing Expanded}
        F_{\sigma\sigma'}(z) &= -\frac{1}{2}\int \mathbf{d^2k}_\perp\sum_{nm}\Big[\expval{\Psi^{(n)}_{-\mathbf{k}_\perp,\downarrow,\sigma}\Psi^{(m)}_{\mathbf{k}_\perp,\uparrow,\sigma'}}\nonumber\\
        & \pm\expval{\Psi^{(n)}_{\mathbf{k}_\perp,\downarrow,\sigma'}\Psi^{(m)}_{-\mathbf{k}_\perp,\uparrow,\sigma}}\Big]\sin(k_nz)\sin(k_mz)
    \end{align}
    We apply a Bogoliubov Transform:
    \begin{align}\label{eq:Final Pairing}
        \Psi^{(n)}_{\mathbf{k}_\perp,s,\sigma} &= \sum_{\alpha}[u^{(n)}_{\alpha,\mathbf{k}_\perp,s,\sigma}\gamma_{\alpha,\mathbf{k}_\perp} + v^{*(n)}_{\alpha,-\mathbf{k}_\perp,s,\sigma}\gamma^\dagger_{\alpha,\mathbf{k}_\perp}]
    \end{align}
    Where $\gamma_{\alpha,\mathbf{k}_\perp}$ ($\gamma^\dagger_{\alpha,\mathbf{k}_\perp}$) is the quasi-particle annihilation (creation) operator for a quasi-particle with energy $E_\alpha$. The pairing amplitude is now given by:
    \begin{align}
        F_{\sigma\sigma'}(z)&= -\frac{1}{2}\sum_{|\zeta_\alpha|\leq\omega_D}\sum_{nm}\int \mathbf{d^2k}_\perp[u^{(n)}_{\alpha,-\mathbf{k}_\perp,\downarrow,\sigma}v^{*(m)}_{\alpha,\mathbf{k}_\perp,\uparrow,\sigma'}\nonumber\\
        &\pm u^{(n)}_{\alpha,\mathbf{k}_\perp,\downarrow,\sigma'}v^{*(m)}_{\alpha,-\mathbf{k}_\perp,\uparrow,\sigma}]sin(k_nz)\sin(k_mz)
    \end{align}
    where $\zeta_\alpha = E_\alpha - \mu$ provides a cutoff for the interaction, and the BCS ground state is the vacuum state of the quasiparticle (i.e. $\expval{\gamma^{\dagger}_{\alpha,\mathbf{k}_\perp}\gamma_{\alpha,\mathbf{k}_\perp}} = 0$). Parameters are chosen such that the Debye window rests above the bottom of the host metallic band and below intersections of the Weyl bands. 
    

    \par
\begin{figure}
        \centering
         \includegraphics[width=0.48\textwidth]{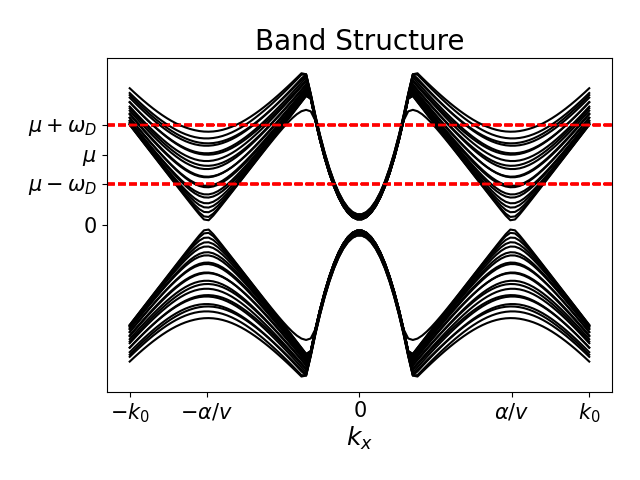}
         \caption{Band structure of the proximitized model in the particle subspace. The Debye window, shown as dashed red lines, is chosen such that there is no overlap between the Weyl and Metallic subspaces. The parameters used are: $N=145$, $E_0=0.05$, $\Delta_0=0.1$, $\omega_D=0.3$, $m_z=3$, $m=2$, $\alpha=2$, $v=1$, and $\mu=0.71$.}
         \label{fig:Band_Structure}
     \end{figure}

 A cut of the noninteracting band structure along $k_y=0$ in the particle subspace is shown in Fig.(\ref{fig:Band_Structure}). With the appropriate choice of parameters and the Debye window (dashed red), the numerical model faithfully approximates a set of Weyl and Metallic bands that couple across the interface. 

\section{Results}

We consider two different architectures. First, we analyze the induced superconductivity in a device with a superconductor placed adjacent to a WSM. The mismatch in band structure results in superconductivity limited to the interface with very little leaking into the WSM. To verify that this is not a result of boundary conditions, we next look at a Superconductor-WSM-Superconductor device. 

\subsection{Superconductor-WSM}\label{scwsm}
\begin{figure*}
    \centering
    \begin{subfigure}[b]{0.48\textwidth}
         \centering
         \includegraphics[width=\textwidth]{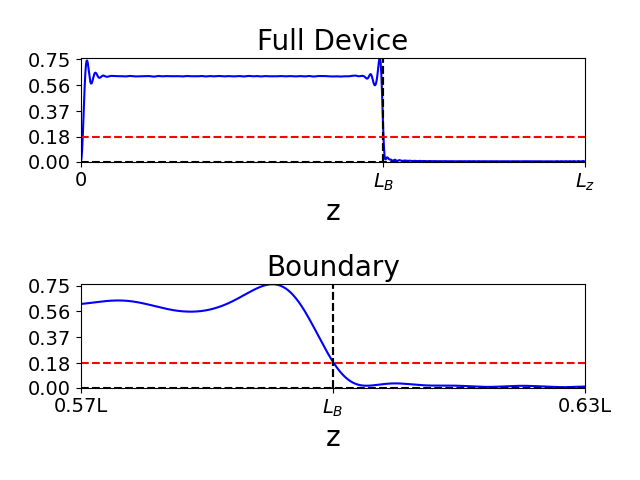}
         \caption{$Re(F_{11})/F_0$}
         \label{fig:Real_11_6}
    \end{subfigure}
    \hfill
    \begin{subfigure}[b]{0.48\textwidth}
         \centering
         \includegraphics[width=\textwidth]{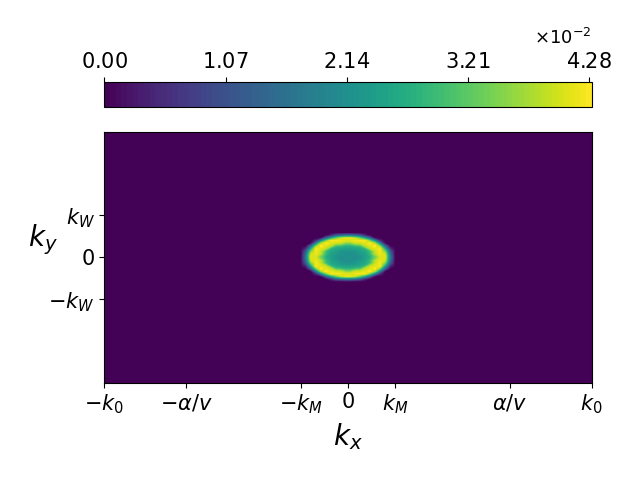
         }
         \caption{$|F_{11}(\mathbf{k}_\perp,z=0.30L)|$}
         \label{fig:RMag_11_SC_6}
    \end{subfigure}
    \hfill
    \begin{subfigure}[b]{0.48\textwidth}
         \centering
         \includegraphics[width=\textwidth]{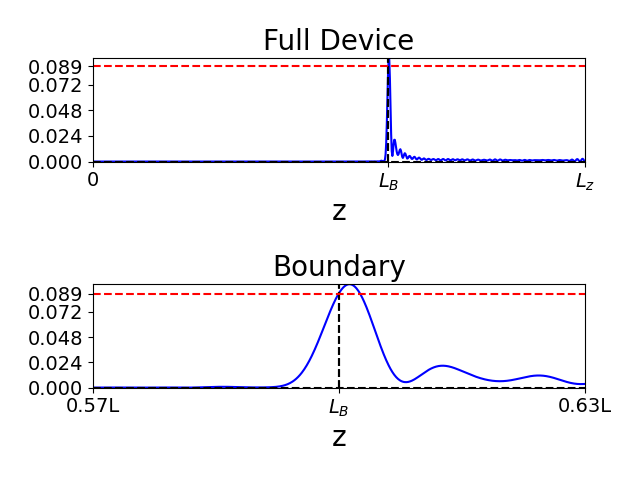}
         \caption{$Re(F_{22})/F_0$}
         \label{fig:Real_22_6}
     \end{subfigure}
    \hfill
    \begin{subfigure}[b]{0.48\textwidth}
         \centering
         \includegraphics[width=\textwidth]{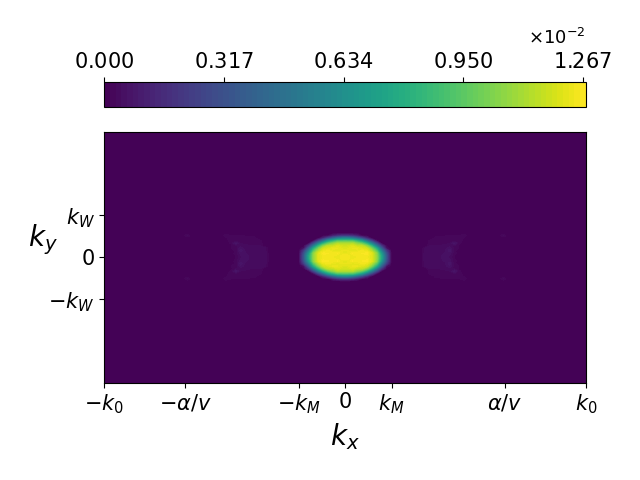}
         \caption{$|F_{11}(\mathbf{k}_\perp,z=0.60L)|$}
         \label{fig:RMag_11_B_6}
    \end{subfigure}
    \hfill
     \begin{subfigure}[b]{0.48\textwidth}
         \centering
         \includegraphics[width=\textwidth]{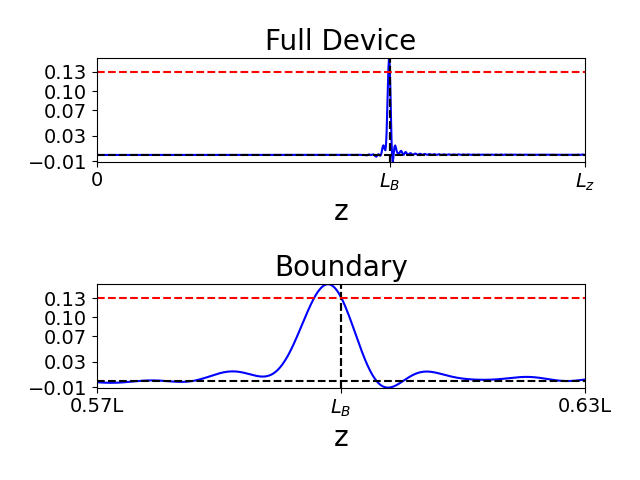}
         \caption{$Re(F_{T})/F_0$}
         \label{fig:Real_T_6}
     \end{subfigure}
     \hfill
    \begin{subfigure}[b]{0.48\textwidth}
         \centering
         \includegraphics[width=\textwidth]{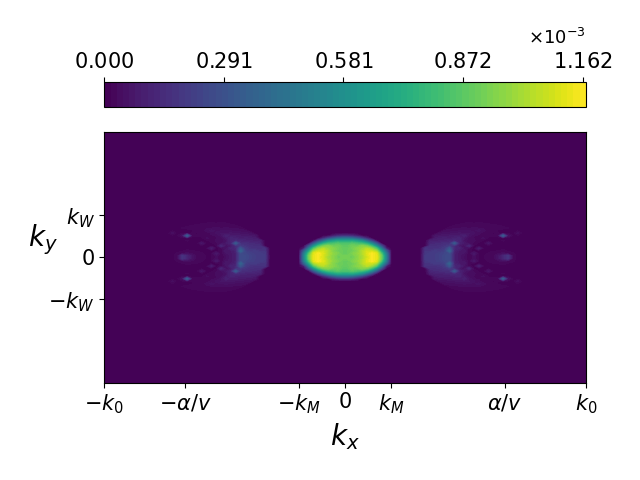}
         \caption{$|F_{11}(\mathbf{k}_\perp,z=0.61L)|$}
         \label{fig:RMag_11_B1_6}
    \end{subfigure}
     
     \caption{(Left) Real component of  (a) $F_{11}/F_0$, (c) $F_{22}/F_0$, and (e) $F_{T}/F_0$ throughout the device (top) and around the boundary (bottom). The pairing amplitude decays sharply near the interface. (b,d,f) Momentum space behavior of $|F_{11}|$ (b) in the center of the host SC, (d) on the interface, and (f) just within the interface on the WSM side. The Weyl nodes are marked at $k_x = \pm\alpha/v$, and the edeges of the metallic (Weyl) debye window are marked at $\pm k_M$ ($\pm k_W$). The boundary is placed at $L_B = 0.6L$, and the parameters used are: $N=145$, $E_0=0.05$, $\Delta_0=0.1$, $\omega_D=0.3$, $m_z=3$, $m=2$, $\alpha=2$, $v=1$, and $\mu=0.71$, $g_{11}=48.53$, and $F_0=6.18\times 10^{-3}$.}
     \label{fig:Pairing_Amplitudes}
    \end{figure*}

Shown in Fig.(\ref{fig:Real_11_6}), Fig.(\ref{fig:Real_22_6}) and Fig.(\ref{fig:Real_T_6}) are the pairing channels $F_{11}, F_{22},$ and $ F_{T}$. The $F_{S}$ channel has been omitted since it is several orders of magnitude smaller than the others. The upper panels cover the entire device while the lower panels focus on the behavior near the interface. The magnitudes are  normalized to $F_{0}=\Delta_{0}/g_{11}$, with $g_{11}\approx 49$ and $F_0\approx 6.1\times10^{-3}$. The broken inversion symmetry implies that a classification in terms of singlets and triplets is not appropriate. This is reflected in the finite amplitude seen in all three pairing channels even though the superconductor is an s-wave spin singlet. However, the mismatch in symmetry and band structure across the interface leads to a significant reduction in amplitude which decays very quickly as one enters the semi-metal. While the peak and oscillatory behavior near the interface are expected from the finite number of Fourier nodes and the step like change in Hamiltonian, the significant drop-off across the interface is a result of disparity between the semi-metallic and metallic behavior of the low energy electronic states. In the appendix, sec.\ref{metal}, we show that, for metallic bands on both sides, the canonical result of a smooth evolution is recovered.

To better characterize the proximity effect, we analyse the momentum dependence of the superconducting gap. Fig.(\ref{fig:RMag_11_SC_6}) shows the form of the pairing amplitude in the middle of the host SC, whereas Fig.(\ref{fig:RMag_11_B_6}) and Fig.(\ref{fig:RMag_11_B1_6}) show the pairing amplitude on and near the interface on the WSM side. Notably, the majority of weight remains near the $\Gamma$ point until one gets well inside the WSM. However, the amplitude has essentially decayed to zero by that point. This suggests that the confinement of the superconducting pairing to the interface on the WSM side of the junction is correlated to the degree to which the metallic electronic states penetrate the WSM. In other words, the wave-functions participating in the superconductivity at and near the interface inside the WSM 
resemble those of the host superconductor.

Similar behavior is observed by both $F_{22}$ and $F_T$. While these results are anticipated by symmetry, the quantitative suppression (by order of magnitude and larger) can only be determined by the detailed analysis presented here.

\subsection{Superconductor-WSM-Superconductor}\label{scwsmsc}
While the surface state of the WSM at the interface with the superconductor is accurately captured above, those at the other end of the device are ignored. Since the induced superconductivity is localized to the region around the interface, this approximation is expected to be valid. To verify this, we next turn to the behavior of a SC-WSM-SC device. To better capture the physics, a greater momentum space resolution is implemented.

\begin{figure*}
    \centering
    \begin{subfigure}[b]{0.48\textwidth}
         \centering
         \includegraphics[width=\textwidth]{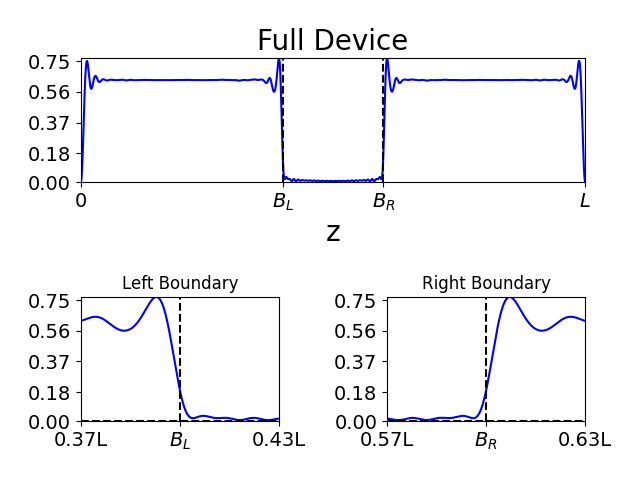}
         \caption{$Re(F_{11})/F_0$}
         \label{fig:Real_11_JJ}
    \end{subfigure}
    \hfill
    \begin{subfigure}[b]{0.48\textwidth}
         \centering
         \includegraphics[width=\textwidth]{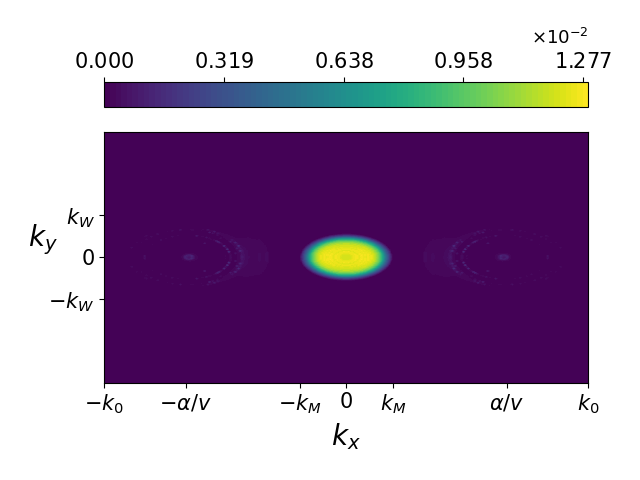}
         \caption{$|F_{11}(\mathbf{k}_\perp,B_L)|$}
         \label{fig:RMag_11_B_JJ}
    \end{subfigure}
    \hfill
    \begin{subfigure}[b]{0.48\textwidth}
         \centering
         \includegraphics[width=\textwidth]{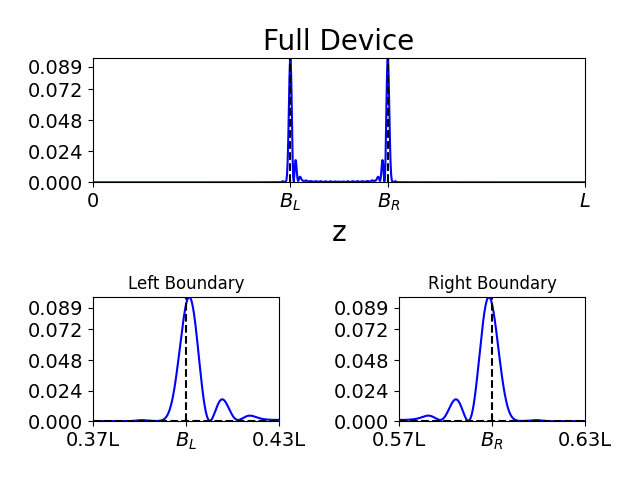}
         \caption{$Re(F_{22})/F_0$}
         \label{fig:Real_22_JJ}
     \end{subfigure}
     \hfill
     \begin{subfigure}[b]{0.48\textwidth}
         \centering
         \includegraphics[width=\textwidth]{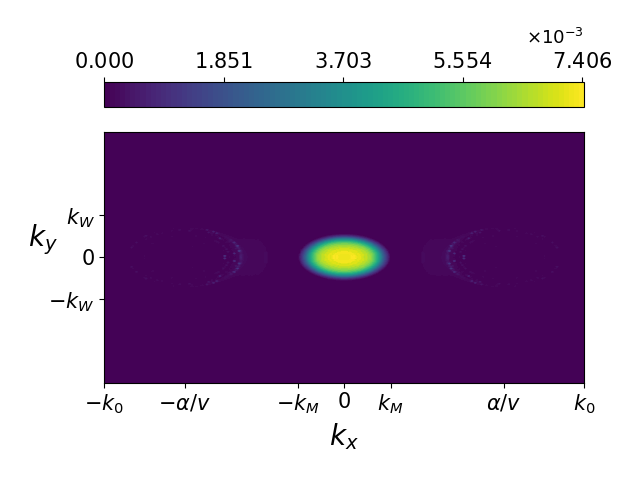}
         \caption{$|F_{22}(\mathbf{k}_\perp,B_L)|$}
         \label{fig:RMag_22_B_JJ}
     \end{subfigure}
     \hfill
     \begin{subfigure}[b]{0.48\textwidth}
         \centering
         \includegraphics[width=\textwidth]{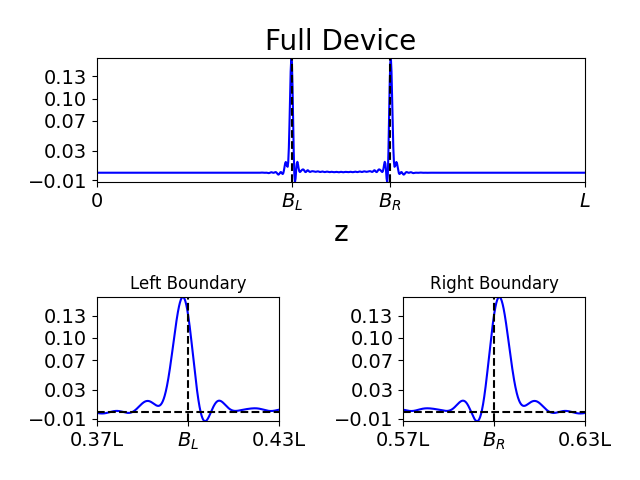}
         \caption{$Re(F_{T})/F_0$}
         \label{fig:Real_T_JJ}
     \end{subfigure}
     \hfill
     \begin{subfigure}[b]{0.48\textwidth}
         \centering
         \includegraphics[width=\textwidth]{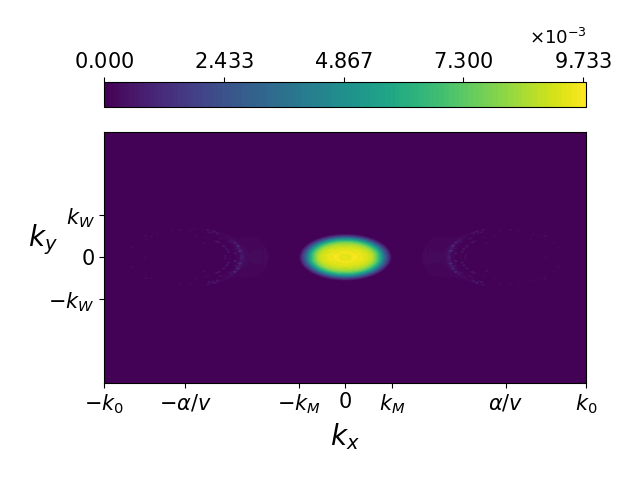}
         \caption{$|F_{T}(\mathbf{k}_\perp,B_L)|$}
         \label{fig:RMag_T_B_JJ}
     \end{subfigure}
     
     \caption{(Left) Real component of (a) $F_{11}/F_0$, (c) $F_{22}/F_0$, and (e) $F_{T}/F_0$ of the two Josephson-Junction. (Right) Momentum space behavior on the left boundary of (b) $F_{11}$, (d) $F_{22}$, and (f)$F_{T}$. The boundaries are placed at $B_L = 0.4L$ and $B_R = 0.6L$, and the parameters used are: $N=145$, $E_0=0.05$, $\Delta_0=0.1$, $\omega_D=0.3$, $m_z=3$, $m=2$, $\alpha=2$, $v=1$, $\mu=0.71$, $g_{11}=48.53$, and $F_0=6.18\times 10^{-3}$.}
     \label{fig:Pairing_Amplitudes_JJ}
    \end{figure*}
        
     Plotted in Fig.(\ref{fig:Real_11_JJ}) is the real component of the pairing mode $F_{11}$ throughout the device, as well as its behavior near the boundaries. The results are in agreement with those in section \ref{scwsm} where induced superconductivity is predominantly in the $F_{11}$ channel confined to the interface. The behaviors of $F_{22}$ and $F_{T}$ are shown in fig.(\ref{fig:Real_22_JJ}) and Fig.(\ref{fig:Real_T_JJ}). The latter are finite as expected by the broken inversion symmetry but are much weaker as compared to the $F_{11}$ channel. 

     The momentum space dependence at the boundary for $F_{11}, F_{22}$ and $F_{T}$ are shown in Fig.(\ref{fig:RMag_11_B_JJ}), fig.(\ref{fig:RMag_22_B_JJ}) and Fig.(\ref{fig:RMag_T_B_JJ}) respectively. As in the single interface case the majority of the weight remains near the $\Gamma$ point in all three channels reflecting the very wek coupling to the Weyl nodes.
     
\begin{figure*}
    \centering
     \begin{subfigure}[b]{0.48\textwidth}
         \centering
         \includegraphics[width=\textwidth]{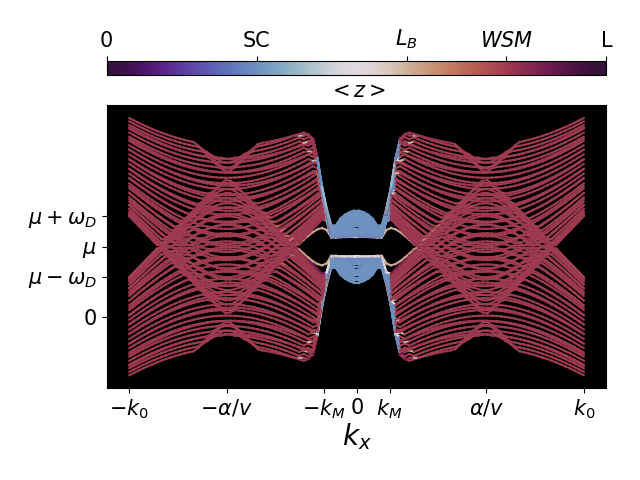}
         \caption{$v=0.5v^c_\ell$}
         \label{fig:Bands_Weighted_50_AVG}
    \end{subfigure}
    \begin{subfigure}[b]{0.48\textwidth}
         \centering
         \includegraphics[width=\textwidth]{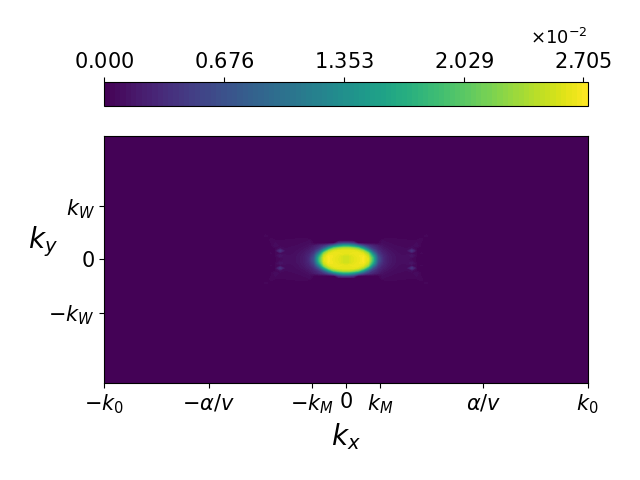}
         \caption{$v=0.5v^c_\ell$}
         \label{fig:Mag_11_B_50}
     \end{subfigure}
    \begin{subfigure}[b]{0.48\textwidth}
         \centering
         \includegraphics[width=\textwidth]{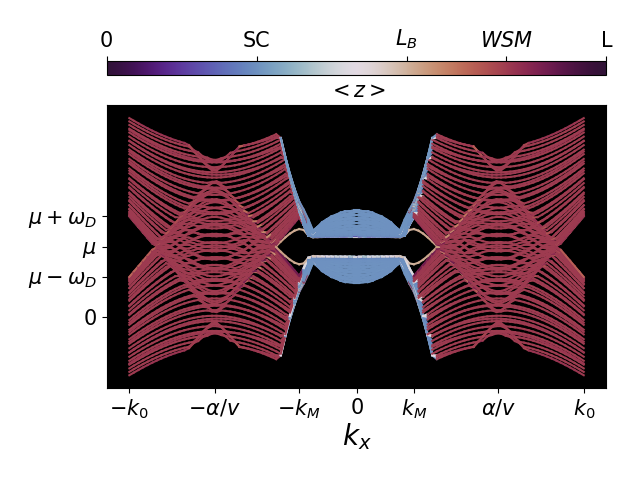}
         \caption{$v=v^c_u$}
         \label{fig:Bands_Weighted_U100_AVG}
     \end{subfigure}
     \begin{subfigure}[b]{0.48\textwidth}
         \centering
         \includegraphics[width=\textwidth]{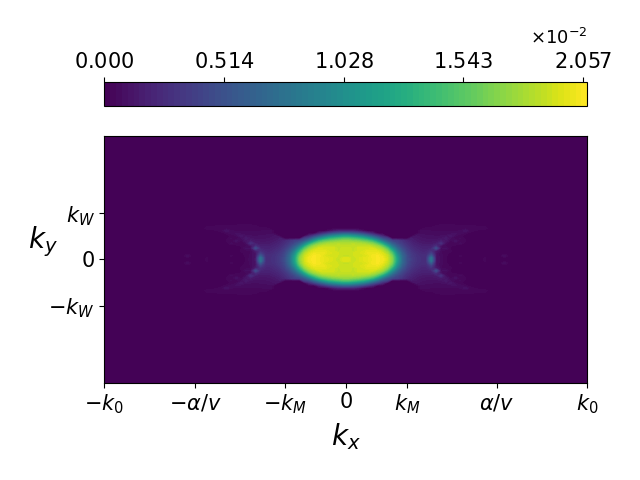}
         \caption{$v=v^c_u$}
         \label{fig:Mag_11_B_U100}
     \end{subfigure}
     \begin{subfigure}[b]{0.48\textwidth}
         \centering
         \includegraphics[width=\textwidth]{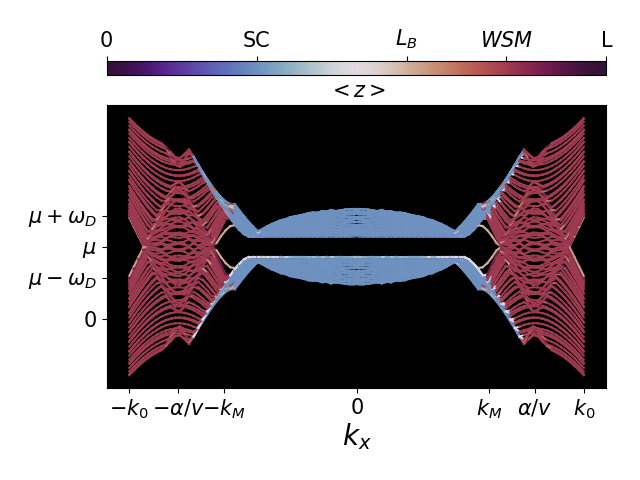}
         \caption{$v=4.0v^c_u$}
         \label{fig:Bands_Weighted_U400_AVG}
     \end{subfigure}
     \begin{subfigure}[b]{0.48\textwidth}
         \centering
         \includegraphics[width=\textwidth]{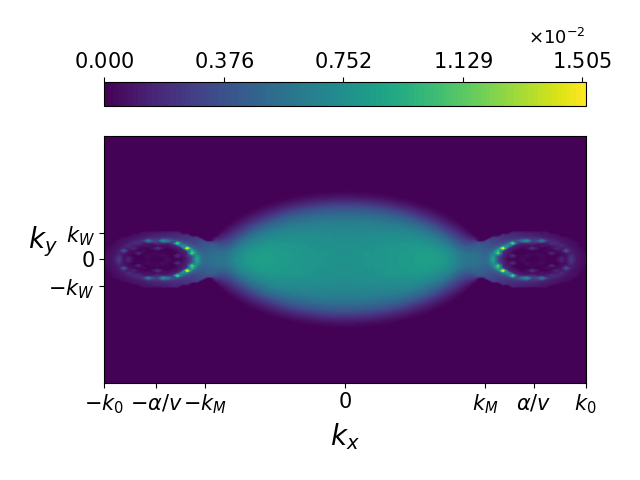}
         \caption{$v=4.0v^c_u$}
         \label{fig:Mag_11_B_U400}
     \end{subfigure}
     \caption{(Left Column) BdG energy bands of the single SC-WSM system for (a) $v=0.5v^c_\ell$, (b)$v=v^c_u$, and (c)$v=4.0v^c_u$. The bands are color weighted by the average of their wave function over the length of the device. 
     (Right Column) Corresponding momentum space distribution for $F_{11}$ at the interface. The model parameters used are: $N=146$, $E_0=0.05$, $\Delta_0=0.1$, $\omega_D=0.3$, $m_z=2$, $m=2$, $\mu=0.7$, $g_{11}=48.9$, and $F_0=6.13\times 10^{-3}$, with $\alpha$ adjusted based on $v$.}
     \label{fig:Boundary_Bands}
\end{figure*} 

\subsection{Velocity mismatch across the interface}
An important determinant of the coupling across the interface is the mismatch in the perpendicular velocity between states of the host superconductor ($v^{sc}_z$) and the WSM (v$^{w}_z$). To understand its impact, we vary the Weyl velocity $v$ and adjust $\alpha$ to keep the Fermi-surfaces separate; all other parameters are fixed. Two limiting values of $v$ are (1) $v^c_\ell$, below which the two systems share no states with similar energy and velocity, and (2) $v^c_u$, above which there are states for which the two systems have the same energy and velocity. The simulation in sections \ref{scwsm} and \ref{scwsmsc} have a Weyl velocity of $v = 0.89 v^c_\ell$, which suggests that the states near the chemical potential on the two sides of the interface have very different velocities. The derivation of these values is given in the appendix (see sec.\ref{critv}).
\par Shown in Fig.(\ref{fig:Boundary_Bands}) are plots of the energy bands for the BdG equations with $k_y=0$, along with the corresponding momentum space distribution of $F_{11}$ at the interface, for  $v=0.5v^c_\ell$, $v=v^c_u$, and $v=4.0v^c_u$. The energy band plots have been color weighted by the average of their wave function over the device.  To ensure that the Fermi surfaces of the two systems remain well separated, we adjust $\alpha$ such that the two band structures still meet at $E=\mu+\omega_D$. Three distinct band structures are observed: (1) Metallic like bands that average to the center of the SC at $0.3L$ (light blue), (2) Weyl like bands that average to the center of the WSM at $0.8L$ (light maroon), and (3) Edge states bridging the two band structures that average to the interface at $0.6L$ (light brown). We find that, below $v^c_\ell$, the pairing function is mostly confined to the host superconductor and does not couple to the Weyl or Edge states; this is reflected in the form of $F_{11}(\mathbf{k}_\perp)$ as a function of $z$. As $v$ is increased to $v^c_u$, the edge states and the pairing function are able to weakly couple to the Weyl nodes. Finally, at $v=4.0v^c_u$, the edge  states and pairing functions are more evenly distributed between the Weyl nodes and $\Gamma$ point.
\par
\begin{figure}[h!]
    \centering
    \begin{subfigure}[b]{0.45\textwidth}
         \centering
         \includegraphics[width=\textwidth]{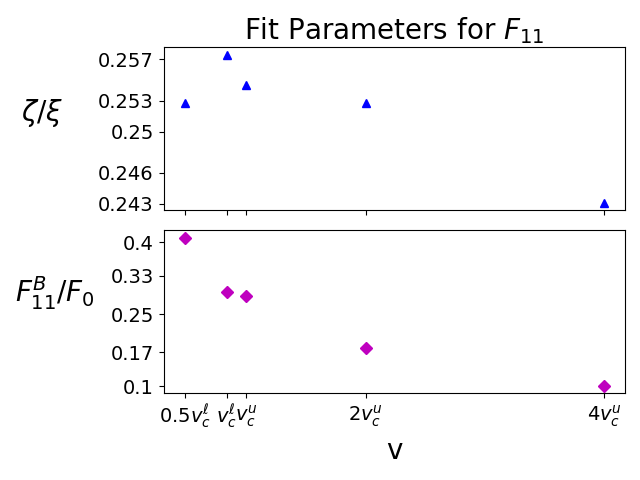}
         \label{fig:Params_11_6}
    \end{subfigure}
    \begin{subfigure}[b]{0.45\textwidth}
         \centering
         \includegraphics[width=\textwidth]{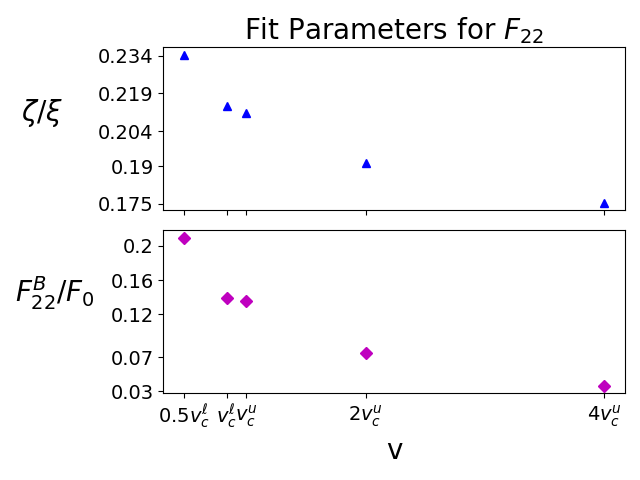}
         \label{fig:Params_22_6}
     \end{subfigure}
     \begin{subfigure}[b]{0.45\textwidth}
         \centering
         \includegraphics[width=\textwidth]{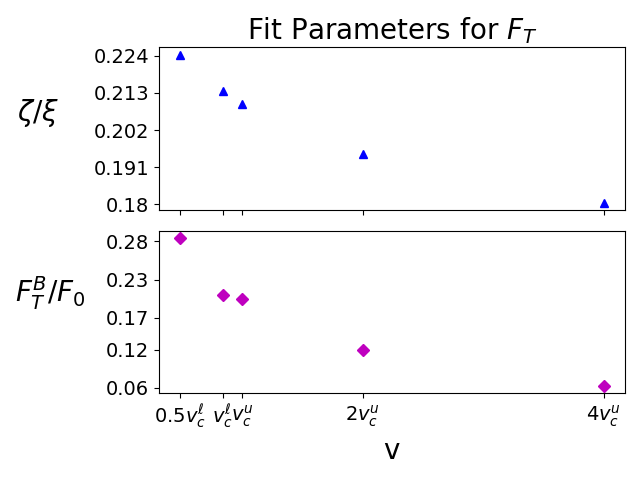}
         \label{fig:Params_T_6}
     \end{subfigure}
     
     \caption{Least squares fit calculation of (Blue) Decay length and (Purple) Interface Amplitude as $v$ is increased for (top) $F_{11}$, (middle) $F_{22}$, and (bottom) $F_T$. $\xi=0.04L$ for the chosen parameters of the simulation. As the velocity mismatch becomes smaller, both the amplitude at the boundary and the coherence length inside the WSM decrease.}
     \label{fig:Parameter_Fits}
\end{figure}

 To better understand the behavior of the pairing modes as $v$ is increased, we fit the real component of each mode in real space to extract the penetration depth $\zeta$ and  the paring amplitude at the interface. These values are plotted and compared to the Cooper pair size $\xi=2m_zk_F/(\pi\Delta_0)$ and the initial pairing amplitude strength $F_0$ in Fig.(\ref{fig:Parameter_Fits}). For a clean superconductor the coherence length is $0.74\xi$. As $v$ is increased, and the pairing amplitudes couple more with the Weyl physics, the amplitude of the pairing modes at the interface and the decay length decrease. This suggests that the mismatch in band structure and loss of inversion symmetry are antagonistic to proximal superconductivity. Even when states with similar velocities and energies exist at the interface the overlap of wave-functions is not sufficient to induce superconductivity well inside the WSM.

\section{Discussion}
A promising architecture often proposed to realize unconventionally, and potentially topological, superconductivity is proximal coupling of an s-wave superconductor to a materials such as WSMs, topological insulators and other unconventional systems. Theoretical modeling providing support to these approaches employ tunneling models across the interface where the parameters are phenomenological inputs. Of interest for experimental implementation are design principles which inform on an optimal choice of material properties to achieve proximal superconductivity. This study elucidates the effects of proximitized superconductivity in an architecture without assuming new physics at the interface beyond quantum tunneling. This is achieved by a numerical calculation of the electronic wave functions and, their correlations, by expanding the respective Hamiltonians in a common Fourier basis.

\par Our simulations show that the degree to which the superconductivity and Weyl physics couple is dependent on mismatches in electronic velocity normal to the interface. The two systems are only able to sufficiently couple once the Weyl velocity $v$ reaches some minimum value $v_c^u$; however, all three pairing channels show a negative correlation between the Weyl velocity and their respective decay length and interface amplitude. This suggests that the induced pairing is unable to penetrate far into the bulk of the WSM. Within a continuum model, with quantum tunneling across the interface, predominantly surface superconducting state is induced by proximity. In other words ensuring continuity of wave-function and probability current at a sharp boundary separating two regions is not enough. Other treatments implement the same boundary assuming tunneling \cite{mcmillan} across the interface but cannot capture the decay of the amplitude in the superconductor. Additional physics involving electronic states near the boundary is needed to induce superconductivity inside the bulk of the WSM. These can be implemented by adding an interface potential or using an alternative approach based on transmission/reflection coefficients \cite{landauer, Buttiker}. Determining the boundary conditions that allow for efficient proximity effect in Weyl semi-metals is an interesting next step and beyond the scope of this work.

\par The momentum space pairings reveal higher weight near the $\Gamma$ point while the edge state is distributed around the Weyl Nodes. The inability of the pairing amplitude to penetrate into the bulk of the WSM likely stems from a mismatch in the momentum of their low energy physics. Future simulations for topologically non-trivial systems with low energy physics more closely aligned in momentum space are planned to verify these conclusions. Of particular interest are architectures consisting of the superconducting Transition Metal Dichalcogenide (TMDC) NbSe$_2$ in contact with another TMDC.\\

\section*{Acknowledgements}
This work used Bridges-2 at Pittsburgh Supercomputing Center through allocation DMR 200086 and PHY220045 from the Advanced Cyberinfrastructure Coordination Ecosystem: Services \& Support (ACCESS) program, which is supported by National Science Foundation grants 2138259, 2138286, 2138307, 2137603, and 2138296 \cite{10.1145/3437359.3465593}. 
\par This work also used the Extreme Science and Engineering Discovery Environment (XSEDE), which is supported by National Science Foundation grant number ACI-1548562. Specifically, it used the Bridges-2 system, which is supported by NSF award number ACI-1928147, at the Pittsburgh Supercomputing Center (PSC)\cite{6866038}.

\bibliography{References}
\clearpage
\newpage
\section*{Appendix}
\subsection{Derivation of Critical Velocities}\label{critv}
In this section we determine the critical velocity $v^c_\ell$, below which the two band structures never have identical z-velocity, and $v^c_u$, above which the two band structures are guaranteed to have states with identical z-velocity. The velocity $v_z = \partial_{k_z}E(\vk)$ for the metallic and Weyl band structures. They are:
\begin{align}
    v_z^W &= \frac{2m_zk_z(m_zk_z^2-m)}{\sqrt{v^2(k_x\pm\alpha/v)^2 + v^2k_y^2 + (m_zk_z^2-m)^2}}\label{eq:v_zW}\\
    v_z^M &= 2m_zk_z\label{eq:v_zM}
\end{align}
It will prove convenient to write the ratio of these two velocities, $R\equiv v_z^W/v_z^M$, in terms of the band energy $E$, the Weyl wave vector magnitude $k^2_W = (k_x\pm\alpha/v)^2 + k_y^2$, and the metallic wave vector magnitude $k^2_M = k_x^2 + k_y^2$:
\begin{align}
    R &= \frac{\sqrt{E^2-v^2k^2_W}}{E} \, \sqrt{\frac{\sqrt{E^2 - v^2k^2_W}+m}{E-E_0 - m_zk^2_M}}\label{eq:Ratio}
\end{align}
For $k_W=k_M\equiv k$ with functions $f(k)=\sqrt{E^2-v^2k^2_W}$ and $g(k)=E-E_0 - m_zk^2_M$ Eq.(\ref{eq:Ratio}) is:
\begin{align}
    R(k,E) &= \frac{f(k)}{E}\sqrt{\frac{f(k)+m}{g(k)}}\label{eq:simplified_ratio}
\end{align}
We seek a condition on our parameters that will either forbid or allow $R(k,E) = 1$. A local extreme exists at $k=0$ which has the value:
\begin{align*}
    R(0,E) = \sqrt{\frac{E+m}{E-E_0}}\equiv h_0 \geq 1
\end{align*}
Determining $v^c_\ell$ is equivalent to finding $v$ for which the concavity of $R(0,E)$ changes sign:
\begin{align}
        R''(0) &= \frac{m_zh_0}{g_0} - \frac{v^2h_0}{f^2_0}\bigg[\frac{2g_0h^2_0 + f_0}{2g_0h^2_0}\bigg]\nonumber\\
        \nonumber\\
        \implies v^\ell_c &= \sqrt{\frac{E+m}{E-E_0}}\sqrt{\frac{2m_zE^2}{3E+2m}}
    \end{align}

\begin{figure}[hbt!]
     \centering
     \includegraphics[width=0.48\textwidth]{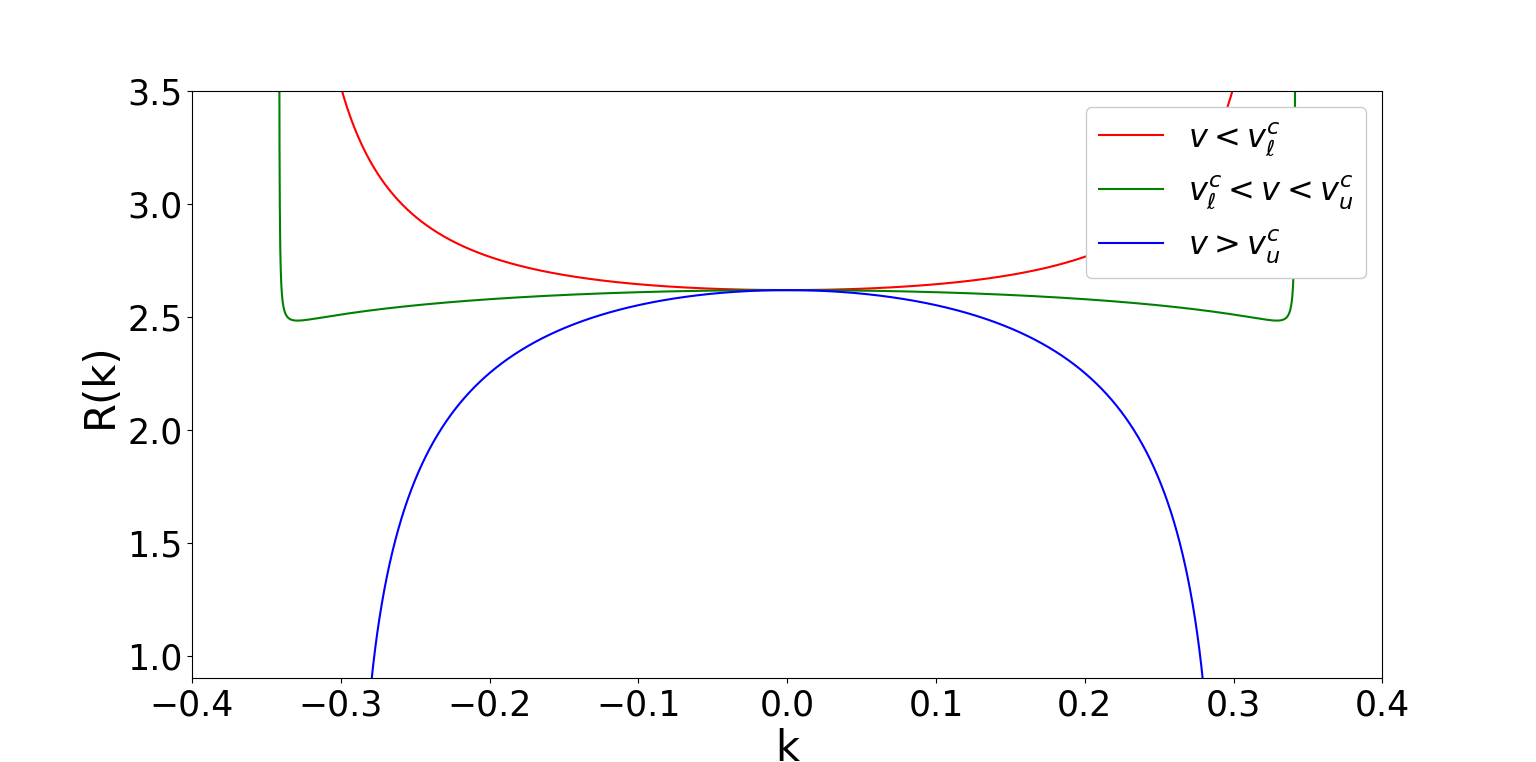}
     \caption{Plot of $R(k)$ for (red) $v<v^c_\ell$, (green) $v^c_\ell < v < v^c_u$, and (blue) $v>v^c_u$ using the same parameters as our simulations and $E=\mu-\omega_D = 0.4$. Only $v>v^c_u$ guarantees a ratio of one.}
     \label{fig:Ratio_Plot}
\end{figure}

For $v<v^c_\ell$, $h_0$ is a global minimum, and thus the electronic velocities are never equal. However, $v>v^c_\ell$ is not enough to guarantee equal velocities, as seen in Fig.(\ref{fig:Ratio_Plot}). We denote $k^W_c$ ($k^M_c$) to be the root of $f(k)$ ($g(k)$). When $k^W_c < k^M_c$, the ratio function diverges before it can reach one; thus, the value $v^c_u$ is obtained when the two roots are equivalent:
\begin{align}
    v^c_u &= \sqrt{\frac{m_zE^2}{E-E_0}} = v^c_\ell\sqrt{\frac{3E+2m}{2E+2m}}
\end{align}
\newpage

For $v^c_\ell<v<v^c_u$, it is still possible to have $R(k,E)=1$ for some value of $k$; in practice, however, this window is quite small and does not guarantee a ratio of one. 
\subsection{Metallic Simulations}\label{metal}
To demonstrate that the numerical approach faithfully accounts for proximal superconductivity, we replace the Weyl Hamiltonian with a metallic Hamiltonian (Eq.(\ref{eq:Metallic_Model})). We study three cases.

\begin{figure*}
    \centering
    \begin{subfigure}[b]{0.48\textwidth}
         \centering
         \includegraphics[width=\textwidth]{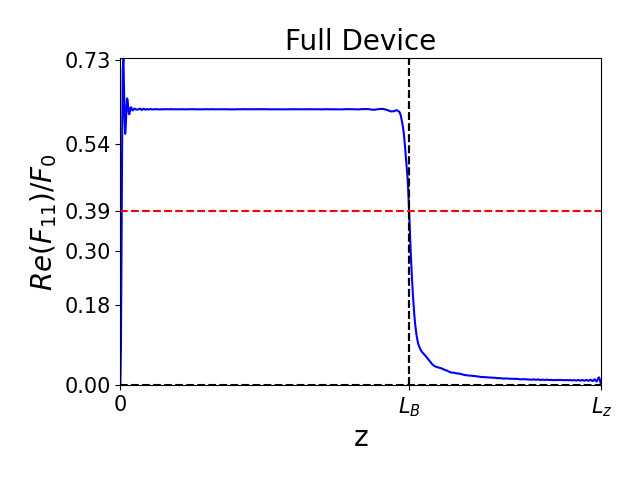}
         \caption{$\alpha=0, E_0=0.05$}
         \label{fig:Pairing_Metal_BIS_0_1}
    \end{subfigure}
    \hfill
    \begin{subfigure}[b]{0.48\textwidth}
         \centering
         \includegraphics[width=\textwidth]{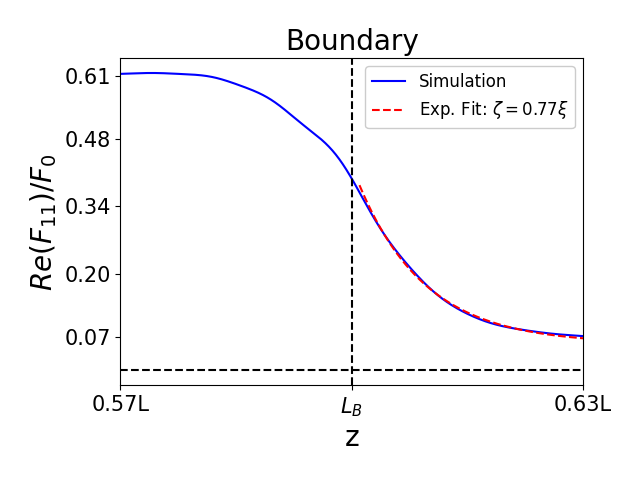}
         \caption{$\alpha=0, E_0=0.05$}
         \label{fig:Pairing_Metal_BIS_0_2}
    \end{subfigure}
    \hfill
    \begin{subfigure}[b]{0.48\textwidth}
         \centering
         \includegraphics[width=\textwidth]{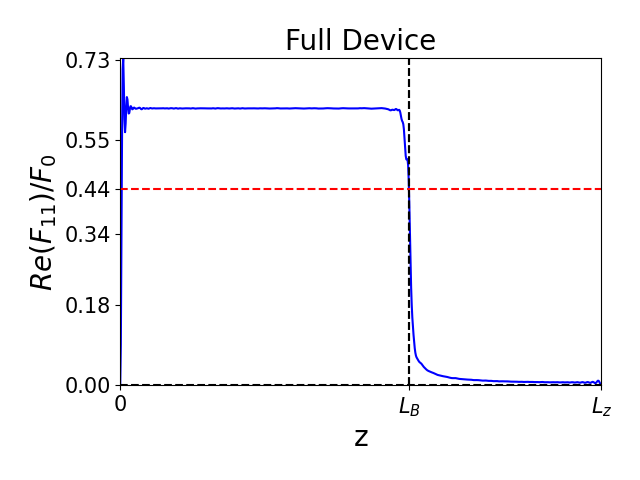}
         \caption{$\alpha=0, E_0=0.3$}
         \label{fig:Pairing_Metal_Shifted_1}
    \end{subfigure}
    \hfill
    \begin{subfigure}[b]{0.48\textwidth}
         \centering
         \includegraphics[width=\textwidth]{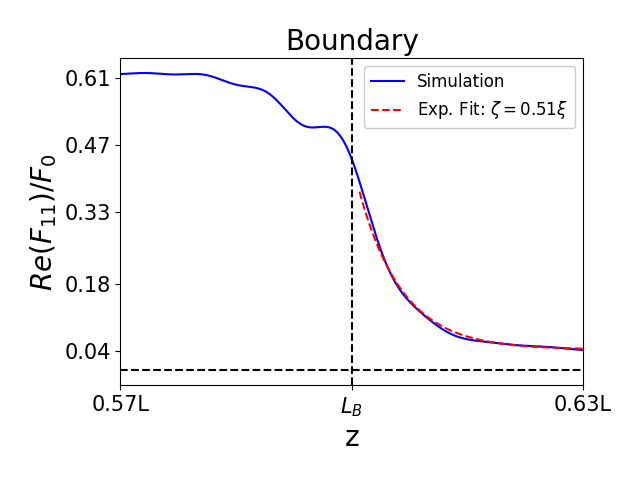}
         \caption{$\alpha=0, E_0=0.3$}
         \label{fig:Pairing_Metal_Shifted_2}
    \end{subfigure}
    \hfill
    \begin{subfigure}[b]{0.48\textwidth}
         \centering
         \includegraphics[width=\textwidth]{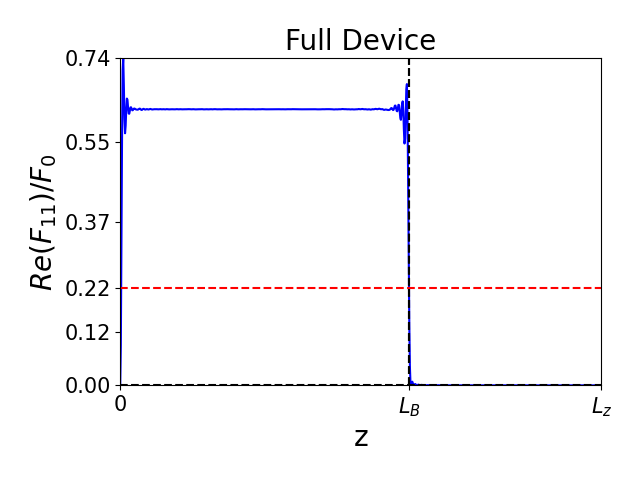}
         \caption{$\alpha=2, E_0=0.05$}
         \label{fig:Pairing_Metal_BIS_2_1}
    \end{subfigure}
    \hfill
    \begin{subfigure}[b]{0.48\textwidth}
         \centering
         \includegraphics[width=\textwidth]{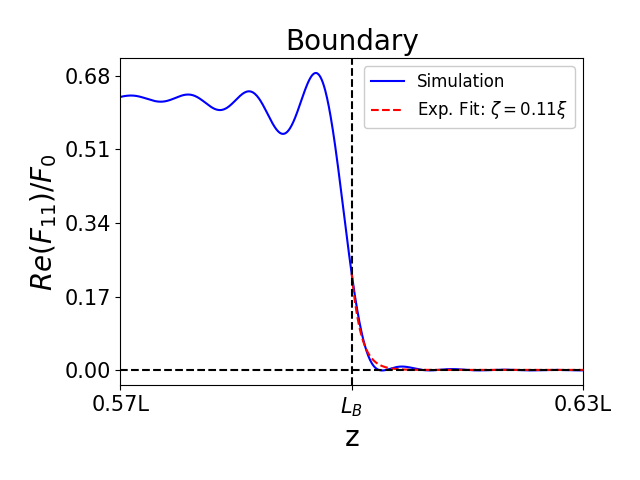}
         \caption{$\alpha=2, E_0=0.05$}
         \label{fig:Pairing_Metal_BIS_2_2}
    \end{subfigure}
    \hfill
     
     \caption{Real component of $F_{11}$ for a metallic model and host superconductor (left) throughout the device and (right) around the interface, where the parameters are varied to explore the effect of the sharp interface: (a-b) identical on both sides, (c-d) shifted band with mismatch in Fermi surface, and (e-f) broken inversion symmetry in the metallic side. All simulations use $N=145$ Fourier modes. $\xi=0.01L$ for the parameters of the simulation.}
     \label{fig:Pairing_Metal}
\end{figure*}

\begin{itemize}

\item The parameters of the Hamiltonian are identical on both sides of the interface. This is the standard model of N-SC junction. In Fig.(\ref{fig:Pairing_Metal_BIS_0_2}), we recover the smooth evolution from the SC to the metal without any oscillatory behavior. Note that for a clean superconductor the coherence length is 0.74$\xi$ which is quantitively consistent with the data.

\item To introduce a mismatch at the interface we introduce a relative shift of the bands. The net effect is to introduce mismatch in velocity and density of states at the chemical potential. While an oscillatory behavior is beginning to emerge in Fig.(\ref{fig:Pairing_Metal_Shifted_2}), a sharp drop off or an evanescent behavior is not observed.

\item To test the effect of inversion breaking, we introduce Eq.(\ref{eq:Inv_Break}) to the metallic side of the interface. A sharp change in symmetry across the interface leads to a sharper fall off and the emergence of oscillatory behavior (see Fig.(\ref{fig:Pairing_Metal_BIS_2_2})). 

We estimate the decay length of each case by fitting to an exponential decay. For these simulations, the Cooper pair size $\xi$ is approximately $0.01L$. 

\end{itemize}

\end{document}